\documentclass[prb,amsmath,amssymb,lengthcheck,showpacs,floatfix,groupedaddress,superscriptaddress,longbibliography]{revtex4-1}
\usepackage{graphicx}
\usepackage[english]{babel}
\usepackage{times}
\usepackage{dcolumn}
\usepackage[usenames,dvipsnames]{color}
\usepackage{units}
\usepackage{bm}

\begin{document}

\title{Non-Fermi-liquid behavior in quantum impurity models with
superconducting channels}

\author{Rok \v{Z}itko}

\affiliation{Jo\v{z}ef Stefan Institute, Jamova 39, SI-1000 Ljubljana, Slovenia}
\affiliation{Faculty  of Mathematics and Physics, University of Ljubljana,
Jadranska 19, SI-1000 Ljubljana, Slovenia}

\author{Michele Fabrizio}
\affiliation{International School for
  Advanced Studies (SISSA), and CNR-IOM Democritos, Via Bonomea
    265, I-34136 Trieste, Italy}

\date{\today}

\begin{abstract}
We study how the non-Fermi-liquid nature of the overscreened
multi-channel Kondo impurity model affects the response to a BCS
pairing term that, in the absence of the impurity, opens a gap
$\Delta$.  We find that non-Fermi liquid features do persist even at
finite $\Delta$: the local density of states lacks coherence peaks, the
states in the continuum above the gap are unconventional, and the
boundary entropy is a non-monotonic function of temperature. Even more
surprisingly, we also find that the low-energy spectrum in the
limit $\Delta\to 0$ actually does not correspond to the spectrum strictly at
$\Delta=0$. In particular, the $\Delta\to 0$ ground state is an
orbitally degenerate spin-singlet, while it is an orbital singlet with
a residual spin degeneracy at $\Delta=0$. In addition, there are
fractionalized spin-1/2 sub-gap excitations whose energy in units of
$\Delta$ tends towards a finite and universal value when $\Delta\to
0$; as if the universality of the anomalous power-law exponents that
characterise the overscreened Kondo effect turned into universal
energy ratios when the scale invariance is broken by $\Delta\not=0$.  This
intriguing phenomenon can be explained
by the renormalisation flow towards the overscreened fixed point and
the gap cutting off the orthogonality catastrophe singularities.  
\end{abstract}


\maketitle

\newcommand{\vc}[1]{{\mathbf{#1}}}
\renewcommand{\Im}{\mathrm{Im}}
\renewcommand{\Re}{\mathrm{Re}}

\newcommand{\expv}[1]{\langle #1 \rangle}
\newcommand{\ket}[1]{| #1 \rangle}
\newcommand{\Tr}{\mathrm{Tr}}

The density of states (DOS) $\rho(\omega)$ of a conventional BCS
$s$-wave superconductor has a gap $\Delta$ and coherence peaks above,
$\rho(\omega) = \rho_0 \Re\, \omega/\sqrt{\omega^2-\Delta^2}$, where
$\rho_0$ is the normal-state DOS. This remains true also in the
presence of disorder and impurities that maintain time-reversal
invariance \cite{anderson1959sc}. By contrast, magnetic impurities may
induce additional states inside the gap by binding Bogoliubov
quasiparticles through the exchange coupling $J$
\cite{shiba1968,sakurai1970,zittartz1968}. The bound-state energies
depend on the interplay of Kondo screening, superconducting proximity
effect, and spin-orbit coupling
\cite{dqdscaniso,moca2008,balatsky2006}, and are measurable in hybrid
superconductor-semiconductor nanostructures
\cite{pillet2010,Deacon:2010jn,maurand2012} and adsorbed magnetic
atoms or molecules \cite{ji2008,franke2011,Hatter:2016kg}. In the
limit of small gap, $\Delta \to 0$, the sub-gap states induced by
impurities that are Kondo screened (effectively non-magnetic) or
underscreened (with a residual local moment decoupled from the rest of
the system) move towards the gap edges and merge with the coherence
peaks, because the weak superconducting pairing perturbs a system that
was formerly in a (regular or singular) local Fermi liquid (FL) state
\cite{nozieres1974,koller2005,mehta2005}. The effect of the impurity
on the bulk electrons is thus fully accounted by the quasiparticle
scattering phase shifts resulting from the Kondo effect ($\pi/2$ in
the deep Kondo limit) \cite{hewson}.
There is, however, a further class of quantum impurities that are
Kondo overscreened because the number of screening channels $k$
exceeds twice the impurity spin $2S$
\cite{mattis1968,nozieres1980,cragg1980}.
Such overcompensation has been experimentally demonstrated for the
two-channel Kondo model ($k=2$, $S=1/2$) in artificial semiconductor
quantum dot devices \cite{oreg2003,potok2007,keller2015,iftikhar2015}.
The resulting states are non-Fermi liquids (NFL)
\cite{nozieres1980,affleck2005} with excitation spectra that deviate
significantly from the FL paradigm, but can be still described in terms of appropriate boundary conformal field theories 
\cite{affleck1991over,emery1992,coleman1995,maldacena1997}.
When a small gap is opened in the contacts of such systems, the
superconducting state thus emerges out of a non-Fermi liquid. Two
related questions arise: 1) What is the nature of the excitations
forming the continuum above the gap? 2) Can the sub-gap states be
interpreted as bound states of NFL excitations?

In this work the problem is studied using the numerical
renormalization group (NRG) technique
\cite{wilson1975,krishna1980a,bulla2008,satori1992} and analytical
arguments. For the two-channel Kondo (2CK) model \cite{cox1998} 
in the limit $\Delta\to 0$ we find surprisingly that 
the low energy spectrum does not reproduce that at $\Delta=0$:
the ground state (GS) is a doubly
degenerate spin-singlet, while two $S=1/2$ sub-gap Shiba states become degenerate with  
a {\sl universal
dimensionless energy ratio}
\begin{equation}
\label{eq1}
\epsilon^* \equiv E^*/\Delta \approx 0.5983.
\end{equation}
In other words, even in the $\Delta\to0$ limit these bound states do not merge with the
continuum. %
The excitations above the gap have NFL degeneracies and spacing, and
there are no coherence peaks in the impurity DOS. When the NFL regime
is disrupted by breaking the channel degeneracy 
\cite{nozieres1980,pang1991}, the sub-gap states do move toward the gap edge and the coherence peaks are restored when the FL-NFL cross-over scale $T^*$ exceeds $\Delta$.

We consider the Hamiltonian $H=\sum_{i=1}^k J_i\, \vc{s}_i \cdot \vc{S}
+ H_i$, where 
\begin{equation}
\begin{split}
H_i &= \sum_{k\sigma} \,\epsilon_k \,c^\dag_{i,k\sigma} c^{\phantom{\dagger}}_{i,k\sigma}
+ \sum_k \,\left(\Delta \, c^\dag_{i,k\uparrow} c^\dag_{i,k\downarrow} +
\text{H.c.}\right),
\end{split}
\end{equation}
i.e., a $k$-channel Kondo model \cite{nozieres1980} with each channel
described by a BCS mean-field Hamiltonian with fixed $\Delta$. $J_i$
is the exchange coupling, $\vc{s}_i$ is the spin density of
channel-$i$ electrons at the position of the impurity, $\vc{S}$ is the
impurity spin-$S$ operator, and finally $c^\dag_{i,k\sigma}$ creates an electron in
channel-$i$ with momentum $k$, spin $\sigma$,
and energy $\epsilon_k$. The continuum has a flat DOS with
half-bandwidth $D$, i.e., $\rho_0=1/2D$.
The Kondo scale for small exchange coupling
is $T_K \approx \exp(-1/\rho_0 J_\mathrm{avg})$, where
$J_\mathrm{avg}=\sum_i J_i/k$ \cite{kolf2007}. If $J_i$ are non-equal,
there is further relevant scale $T^*$ that grows as a power law with the difference between the 
two largest $J_i$
\cite{cox1998}. For $\Delta=0$, the system has NFL properties for $T^*
< T < T_K$ and crosses over to a FL GS for $T<T^*$. For $\Delta \neq
0$, we use the NRG to compute the finite-size excitation spectra,
thermodynamic properties, and the $T$-matrix spectral function
(impurity DOS).

\begin{figure}
\centering
\includegraphics[clip,width=0.5\textwidth]{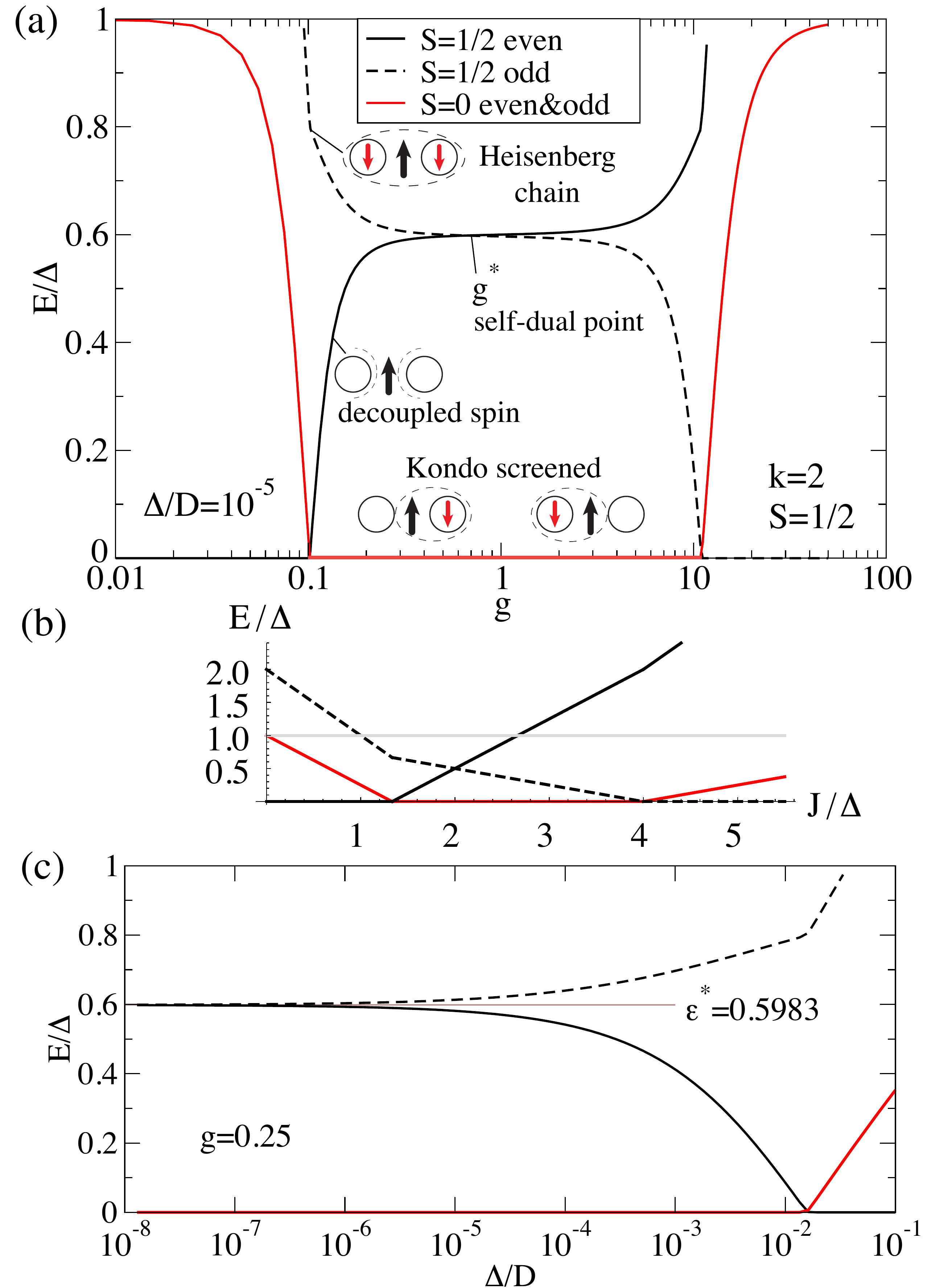}
\caption{(a) Sub-gap many-particle states in the 2CK
model with fixed gap as a function of $g=\rho_0 J$. The energies are
given with respect to the ground-state energy. (b) Eigenstates in the
zero-bandwidth limit. (c) Approach toward the asymptotic universal
spectrum in the zero-gap limit.}
\label{fig1}
\end{figure}

The discrete (sub-gap) part of the excitation spectrum for the 2CK
model with $J_1=J_2=J$ is shown in Fig.~\ref{fig1}(a) at constant gap
$\Delta$ as a function of the dimensionless coupling constant
$g=\rho_0 J$. To better understand the origin of these states, we
introduce a simplified zero-bandwidth model where each screening
channel is represented by a single orbital $f_i$ with pairing $
\big(\Delta\, f^\dag_{i\uparrow} f^\dag_{i\downarrow} +
\text{H.c.}\big)$, and $2\vc{s}_i = \sum_{\alpha\beta}\,
f^\dag_{i\alpha}\,\boldsymbol{\sigma}_{\alpha\beta}\,
f^{\phantom{\dagger}}_{i\beta}$. The lowest four eigenstates, shown in
Fig.~\ref{fig1}(b), are in qualitative correspondence with those of
the full model. The even-parity $S=1/2$ state represents a decoupled
impurity spin. (The parity refers to the channel inversion symmetry.)
This is the GS for low $g$ and corresponds to the local-moment phase
at $\Delta > T_K$. Each spin-singlet state corresponds to the impurity
spin coupled into an $S=0$ state to a singly-occupied orbital, the
other orbital being instead in the configuration
$\big(\ket{0}-\ket{\uparrow\downarrow}\big)/\sqrt{2}$.  There are
evidently two such spin-singlet states, depending on which orbital
screens the impurity. This is actually the doubly degenerate GS for
intermediate values of $g$. Finally, in the large-$g$ limit the GS is
an odd-parity strong-coupling state: both orbitals are singly occupied
and coupled into an odd-parity spin-triplet configuration, which is in
turn coupled to the impurity into a $S=1/2$ state. The low-$g$ and
high-$g$ limits are related through a duality mapping which
interchanges the parity of the $S=1/2$ states and which also occurs in
the $\Delta=0$ model \cite{pang1991,kolf2007}. At $g=g^* \approx 0.7$
these states cross and thereby define the self-dual point. This occurs
at a particular value of the excitation energy $\epsilon^*$ defined in
Eq.~\eqref{eq1}. This value is actually universal: for any $g$ the two
doublet levels converge in the $\Delta \to 0$ limit toward the same
value $\epsilon^*$, see Fig.~\ref{fig1}(c). The self-dual point thus
defines the universal non-trivial fixed point point of the theory in
the small-gap limit, as well as at $\Delta=0$. The approach toward
this limit is a square-root function of $\Delta$:
\begin{equation}
E_{o,e}/\Delta \sim \epsilon^* + c_{o,e}(g) \Delta^{1/2}.
\end{equation}
Magnetic anisotropy $J_\perp \neq J_z$ is irrelevant, as in the normal
state, and does not affect the value of $\epsilon^*$.

\begin{figure}
\centering
\includegraphics[clip,width=0.5\textwidth]{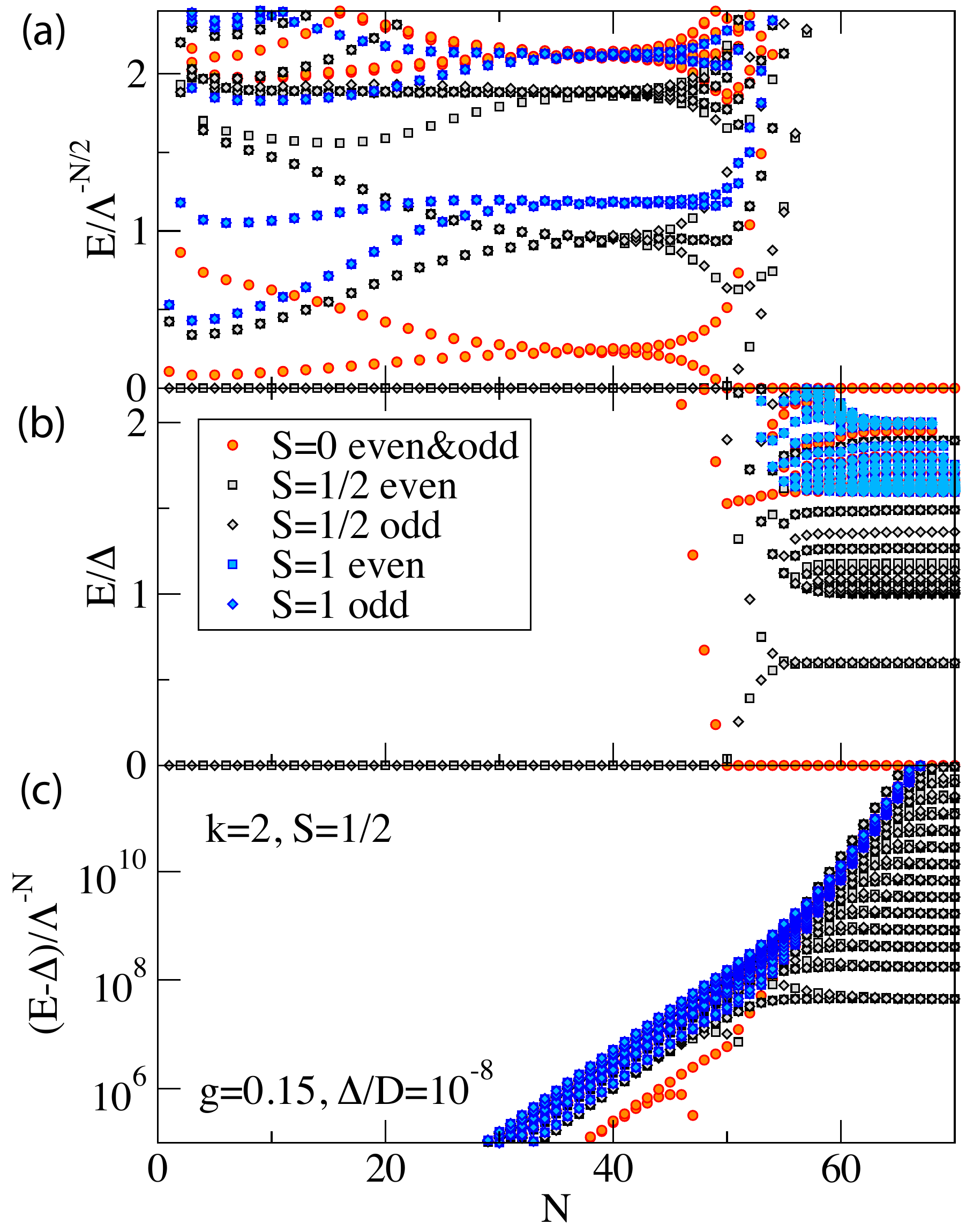}
\caption{Renormalization flow diagram
with different scalings of the energy axis. $N$ is the iteration
number and $\Lambda=2$ is the NRG discretization parameter.}
\label{fig2}
\end{figure}

In Fig.~\ref{fig2} we show the finite-size excitation spectrum (NRG
flow diagram) as a function of the Wilson chain length $N$,
corresponding to the energy scale $\varepsilon_N = \Lambda^{-N/2}$.
Specifically, Fig.~\ref{fig2}(a) reports the energies scaled as $
\epsilon=E/\varepsilon_N$, and demonstrates the cross-over from the
local-moment to the 2CK NFL fixed point for $N \gtrsim 30$, with
characteristic fractional energies $\epsilon_N=0,1/8, 1/2, 5/8, 1,
\ldots$ and degeneracies $2, 4, 10, 12, 26, \ldots$, respectively, 
that reflect the peculiar $SU(2)_2\times SO(5)$ conformal field theory
(CFT) that describes the asymptotic behaviour of the model at
$\Delta=0$ \cite{affleck1991over,maldacena1997,cox1998,supplscnfl}. We observe
that a finite $\Delta$ lowers the $SO(5)$ symmetry down to
$SU(2)\times U(1)$, which corresponds to an $SU(2)_2$ CFT times the
$Z_2$ orbifold of a compactified $c=1$ CFT. The latter allows for a
marginal boundary operator able to split the $SO(5)$ multiplets; for
instance the degeneracy 4 of the 1/8 state into $4\to 2+2$, or the
degeneracy 10 of the 1/2 state into $10\to 2+6+2$ \cite{supplscnfl}. Such splitting is
already evident in Fig.~\ref{fig2}(a) for $40\lesssim N\lesssim 45$.
However, for $N \gtrsim 45$, the BCS gap exceeds $\varepsilon_N$ and
induces flow toward a new fixed point which is better characterized by
scaling the energies as $E/\Delta$, see Fig.~\ref{fig2}(b). The lowest
doublet of the split $\epsilon=1/8$ multiplet becomes the doubly
degenerate spin-singlet GS, while the $\epsilon=0$ $S=1/2$-state and
the lowest $S=1/2$ state of the split $\epsilon=1/2$ multiplet meet
into the $S=1/2$ sub-gap doublet. 
The continuum of excitations for $E > \Delta$, which is dense close to
the gap edge, is best shown scaled as $(E-\Delta)/\Lambda^{-N}$,
Fig.~\ref{fig2}(c). The energies are spaced by the ratio of $\Lambda$,
rather than $\Lambda^2$ as in the gapped single-channel FL case
\cite{hecht2008,supplscnfl}. Such progression results from a combination of FL
states with one channel having $\delta=0$ and the other $\delta=\pi/2$ quasiparticle phase shift, as expected for a GS where the Kondo effect is formed with one channel, the other being decoupled. This does not imply, however, that FL behavior is recovered. The degeneracy of
states above the gap is twice the number for a FL. More remarkably,
when the matrix elements are evaluated to compute the impurity DOS,
there are no coherence peaks (see below). The non-trivial effects are
thus revealed chiefly through the matrix elements, rather than the
energy-level spacing.

\begin{figure}
\centering
\includegraphics[clip,width=0.5\textwidth]{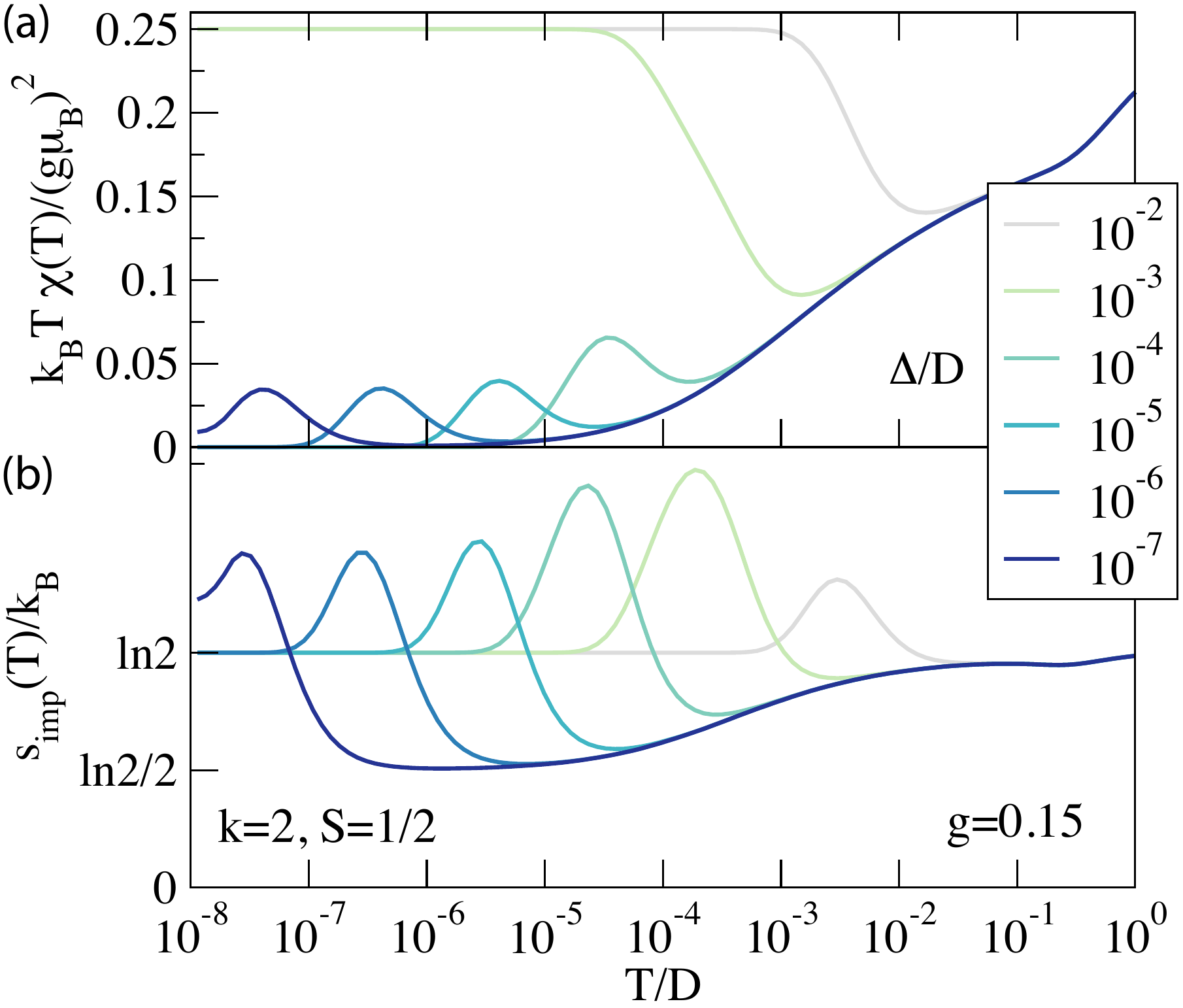}
\caption{Temperature dependence of (a) impurity
magnetic susceptibility and (b) impurity entropy.}
\label{fig3}
\end{figure}

The reshuffling of states when the gap opens leads to peculiar
thermodynamics. In Fig.~\ref{fig3} we plot the impurity magnetic
susceptibility, $\chi_\mathrm{imp}$, and impurity (boundary) entropy
$s_\mathrm{imp}$. For $T \gg \Delta$, the temperature dependence
equals that of the $\Delta=0$ model: the effective local moment goes
to zero and the impurity entropy reaches the $\ln2/2$ plateau. At
$T\sim\Delta$, the effective degeneracy of the impurity-generated
states {\sl increases}, leading to peaks in both $\chi_\mathrm{imp}$
and $s_\mathrm{imp}$. The boundary entropy thus increases from $\sim
\ln2/2$ to $\ln2$. The $g$-theorem \cite{affleck1991prl} does not hold
here because $\Delta$ breaks the conformal invariance.

\begin{figure}
\centering
\includegraphics[clip,width=0.5\textwidth]{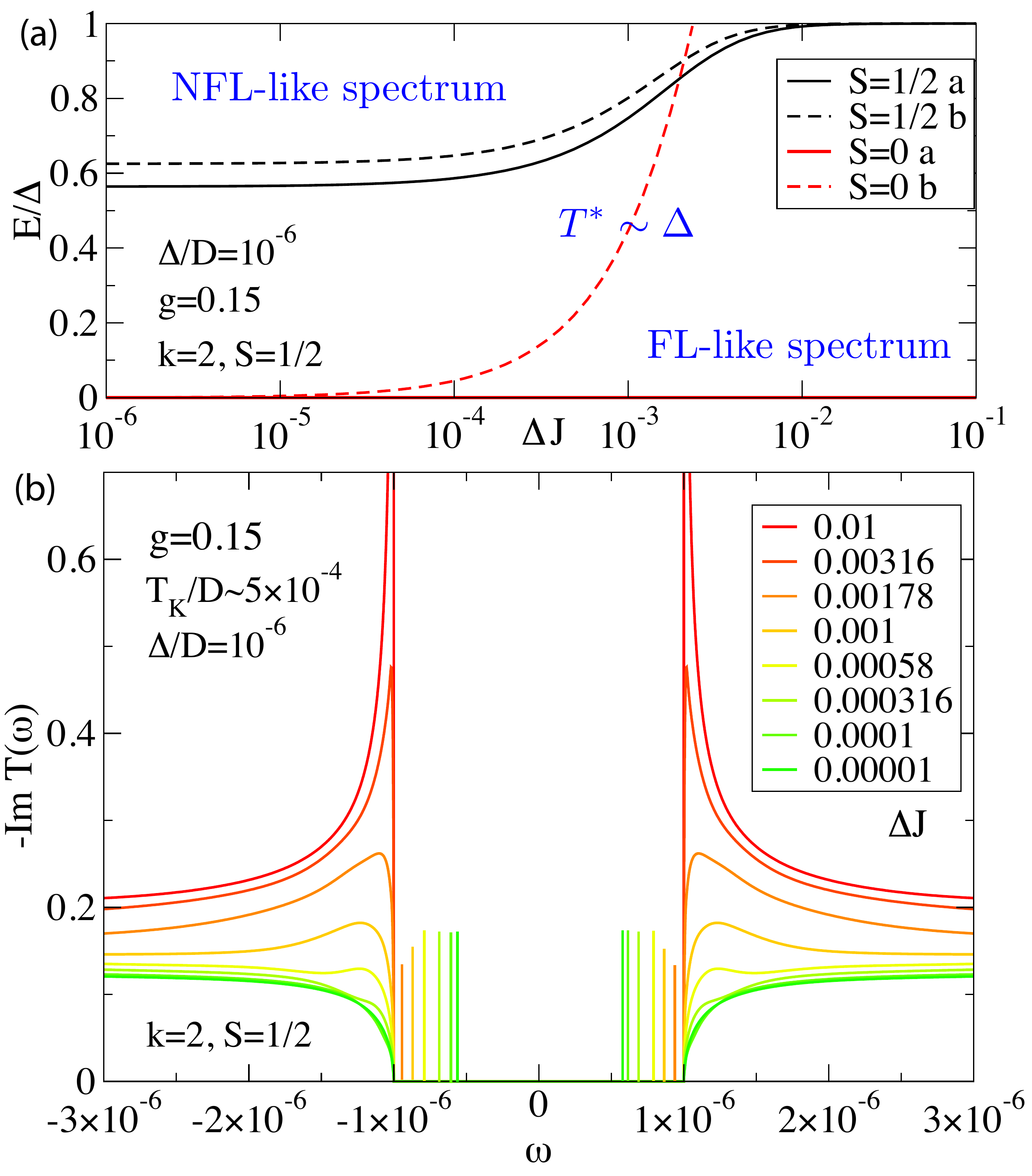}
\caption{Channel symmetry breaking, $J_1 \neq J_2$.
(a) Evolution of the sub-gap states vs. $\Delta J$; the
cross-over point $T^* \sim \Delta$ occurs for $\Delta J \approx
10^{-3}$. (b) Impurity spectral function ($\Im$ part of $T$-matrix)
for a range of $\Delta J$.}
\label{fig4}
\end{figure}

The FL properties are restored when the channel symmetry is broken.
The sub-gap spectrum, shown in Fig.~\ref{fig4}(a) as a function of
$\Delta J=(J_1-J_2)/2$ for constant $J_\mathrm{avg}$ such that $T_K
\gg \Delta$, now depends on the relative value of $\Delta$ and the
NFL-FL crossover scale $T^*$. For small $\Delta J$ such that $T^* \ll
\Delta$, the sub-gap spectrum is ``NFL-like'' with the $S=1/2$ doublet close to $\epsilon^*$. As $\Delta J$ increases, the
degeneracy between the spin-singlet states is lifted. The Kondo state in the dominant channel remains the GS, while the other Kondo state
rapidly rises in energy and enters the continuum. The sub-gap $S=1/2$ doublet asymptotically approaches the gap edge in the large $\Delta J$
limit where $T^* \gg \Delta$, and the spectrum becomes
``FL-like'' with no sub-gap states (because $\Delta \ll T_K$). The
NFL-FL crossover has a characteristic signature also in the impurity
DOS, see Fig.~\ref{fig4}(b): with increasing $\Delta J$, the
$\delta$-peaks move toward the continuum edges and the spectral weight is transferred from the sub-gap region into the continuum to form the coherence peaks characteristic of the FL regime.

We now sketch an analytical argument that explains qualitatively and
to some extent also quantitatively the NRG results; details are
provided as Supplemental Materials \cite{supplscnfl}. At particle-hole
symmetry, the Hamiltonian can be mapped onto a non-superconducting
model in which the pairing potential is transformed into a staggered
one with the same strength $\Delta$, the advantage being that
symmetries and quantum numbers are more transparent. In the same
spirit as Anderson and Yuval\cite{Anderson&Yuval,
Anderson&Yuval&Hamann,1995PhRvL..74.4503F}, we start by the scattering problem at zero
spin-flip exchange, $J_\perp=0$. In this case, if the impurity spin is
up, the spin up and down electrons feel a scattering potential
$V_\uparrow = V$ and $V_\downarrow=-V$, respectively, where $V =
J_z/4$. $V_{\uparrow(\downarrow)}$ changes sign if the impurity spin
is reversed. In the spectrum of the electrons with spin opposite to
the impurity, a channel-degenerate bound state appears whose energy
\begin{equation}
E_\text{bound} = \Delta \;\frac{1-\left( \pi \rho V \right)^2}{1+\left( \pi \rho V \right)^2}\,,
\end{equation}
is located right in the middle of the gap at the self-dual point $(\pi \rho V)^2=1$.
The ground state at $J_\perp=0$ is therefore degenerate. If the
impurity spin is up, there is a spin-down bound state right at the Fermi
level that can be either empty or occupied at no energy cost, hence a
degeneracy two for each channel. Analogously if all spins are
reversed; hence an overall eightfold degeneracy. A finite $J_\perp$
splits this degenerate subspace since it has a finite matrix element
$\propto \big(\rho\Delta\big)^{3/2}$ between states with a single
electron occupying the bound state with the same channel index but
opposite spin. Inclusion of $J_\perp$ at all orders corresponds in the
Anderson and Yuval's approach to an infinite sequence of X-ray edge
problems. However, in this circumstance the edge singularities are cut
off by the gap $\Delta$, so that the effective low-energy model
obtained by integrating out the high-energy states still comprises the
above fourfold degenerate multiplet separated by a finite energy gap
from higher energy states. Such low-energy multiplet is split as
discussed above by an upward renormalised $J_\perp^* \simeq
J_\perp\,\big(\Delta\rho\big)^{-1/2}$ into a lowest energy
channel-degenerate spin-singlet, with singly occupied bound state,
followed by two spin-1/2 states where the bound state is either empty
or doubly occupied and next by a channel-degenerate spin-triplet still
with singly occupied bound state, which however falls in the continuum
above the gap. The energy difference between the two inside-gap
states, the lowest spin-singlet and upper spin-1/2, is estimated as
$(2/\pi) \Delta \approx 0.637 \Delta$, surprisingly close to the
actual value $\epsilon^* \approx 0.598$.

The three-channel Kondo (3CK) models show similar anomalies with
multiple spin-doublet sub-gap states with universal energy ratios
$0.19$ and $0.72$ \cite{supplscnfl}.  The local DOS above the gap is
anomalous, lacking the Fermi-liquid coherence peaks. We speculate that
impurity models with Kondo overscreening generically flow in the
small-gap limit to fixed points with characteristic persistent bound
states of NFL character.

More generally, let us imagine to add the mass term $\Delta$ first at the impurity site. This term does not correspond to any of the relevant boundary operators at the NFL fixed point, nevertheless it lowers the symmetry and allows for a marginal boundary operator. 
If this is the case, we argue that an arbitrarily weak mass term of that kind added at the impurity site as well as in the bulk 
will induce sub-gap states with universal ratios. The predicted
universal spectra are in principle empirically testable in quantum
impurity systems.

\begin{acknowledgments}
R\v{Z} acknowledges the support of the Slovenian Research Agency (ARRS)
under P1-0044 and J1-7259.
\end{acknowledgments}

\bibliography{scnfl}

\begin{thebibliography}{42}%
\makeatletter
\providecommand \@ifxundefined [1]{%
 \@ifx{#1\undefined}
}%
\providecommand \@ifnum [1]{%
 \ifnum #1\expandafter \@firstoftwo
 \else \expandafter \@secondoftwo
 \fi
}%
\providecommand \@ifx [1]{%
 \ifx #1\expandafter \@firstoftwo
 \else \expandafter \@secondoftwo
 \fi
}%
\providecommand \natexlab [1]{#1}%
\providecommand \enquote  [1]{``#1''}%
\providecommand \bibnamefont  [1]{#1}%
\providecommand \bibfnamefont [1]{#1}%
\providecommand \citenamefont [1]{#1}%
\providecommand \href@noop [0]{\@secondoftwo}%
\providecommand \href [0]{\begingroup \@sanitize@url \@href}%
\providecommand \@href[1]{\@@startlink{#1}\@@href}%
\providecommand \@@href[1]{\endgroup#1\@@endlink}%
\providecommand \@sanitize@url [0]{\catcode `\\12\catcode `\$12\catcode
  `\&12\catcode `\#12\catcode `\^12\catcode `\_12\catcode `\%12\relax}%
\providecommand \@@startlink[1]{}%
\providecommand \@@endlink[0]{}%
\providecommand \url  [0]{\begingroup\@sanitize@url \@url }%
\providecommand \@url [1]{\endgroup\@href {#1}{\urlprefix }}%
\providecommand \urlprefix  [0]{URL }%
\providecommand \Eprint [0]{\href }%
\providecommand \doibase [0]{http://dx.doi.org/}%
\providecommand \selectlanguage [0]{\@gobble}%
\providecommand \bibinfo  [0]{\@secondoftwo}%
\providecommand \bibfield  [0]{\@secondoftwo}%
\providecommand \translation [1]{[#1]}%
\providecommand \BibitemOpen [0]{}%
\providecommand \bibitemStop [0]{}%
\providecommand \bibitemNoStop [0]{.\EOS\space}%
\providecommand \EOS [0]{\spacefactor3000\relax}%
\providecommand \BibitemShut  [1]{\csname bibitem#1\endcsname}%
\let\auto@bib@innerbib\@empty
\bibitem [{\citenamefont {Anderson}(1959)}]{anderson1959sc}%
  \BibitemOpen
  \bibfield  {author} {\bibinfo {author} {\bibfnamefont {P.~W.}\ \bibnamefont
  {Anderson}},\ }\bibfield  {title} {\enquote {\bibinfo {title} {Theory of
  dirty superconductors},}\ }\href@noop {} {\bibfield  {journal} {\bibinfo
  {journal} {J. Phys. Chem. Solids}\ }\textbf {\bibinfo {volume} {11}},\
  \bibinfo {pages} {26} (\bibinfo {year} {1959})}\BibitemShut {NoStop}%
\bibitem [{\citenamefont {Shiba}(1968)}]{shiba1968}%
  \BibitemOpen
  \bibfield  {author} {\bibinfo {author} {\bibfnamefont {H.}~\bibnamefont
  {Shiba}},\ }\bibfield  {title} {\enquote {\bibinfo {title} {Classical spins
  in superconductors},}\ }\href@noop {} {\bibfield  {journal} {\bibinfo
  {journal} {Prog. Theor. Phys.}\ }\textbf {\bibinfo {volume} {40}},\ \bibinfo
  {pages} {435} (\bibinfo {year} {1968})}\BibitemShut {NoStop}%
\bibitem [{\citenamefont {Sakurai}(1970)}]{sakurai1970}%
  \BibitemOpen
  \bibfield  {author} {\bibinfo {author} {\bibfnamefont {Akio}\ \bibnamefont
  {Sakurai}},\ }\bibfield  {title} {\enquote {\bibinfo {title} {Comments on
  superconductors with magnetic impurities},}\ }\href@noop {} {\bibfield
  {journal} {\bibinfo  {journal} {Prog. Theor. Phys.}\ }\textbf {\bibinfo
  {volume} {44}},\ \bibinfo {pages} {1472} (\bibinfo {year}
  {1970})}\BibitemShut {NoStop}%
\bibitem [{\citenamefont {Zittartz}\ and\ \citenamefont
  {M\"uller-Hartmann}(1968)}]{zittartz1968}%
  \BibitemOpen
  \bibfield  {author} {\bibinfo {author} {\bibfnamefont {J.}~\bibnamefont
  {Zittartz}}\ and\ \bibinfo {author} {\bibfnamefont {E.}~\bibnamefont
  {M\"uller-Hartmann}},\ }\bibfield  {title} {\enquote {\bibinfo {title}
  {{Green's function theory of the Kondo effect in dilute magnetic alloys}},}\
  }\href@noop {} {\bibfield  {journal} {\bibinfo  {journal} {Z. Physik.}\
  }\textbf {\bibinfo {volume} {212}},\ \bibinfo {pages} {380} (\bibinfo {year}
  {1968})}\BibitemShut {NoStop}%
\bibitem [{\citenamefont {\v{Z}itko}\ \emph {et~al.}(2011)\citenamefont
  {\v{Z}itko}, \citenamefont {Bodensiek},\ and\ \citenamefont
  {Pruschke}}]{dqdscaniso}%
  \BibitemOpen
  \bibfield  {author} {\bibinfo {author} {\bibfnamefont {R.}~\bibnamefont
  {\v{Z}itko}}, \bibinfo {author} {\bibfnamefont {O.}~\bibnamefont
  {Bodensiek}}, \ and\ \bibinfo {author} {\bibfnamefont {Th.}\ \bibnamefont
  {Pruschke}},\ }\bibfield  {title} {\enquote {\bibinfo {title} {Magnetic
  anisotropy effects on quantum impurities in superconducting host},}\
  }\href@noop {} {\bibfield  {journal} {\bibinfo  {journal} {Phys. Rev. B}\
  }\textbf {\bibinfo {volume} {83}},\ \bibinfo {pages} {054512} (\bibinfo
  {year} {2011})}\BibitemShut {NoStop}%
\bibitem [{\citenamefont {Moca}\ \emph {et~al.}(2008)\citenamefont {Moca},
  \citenamefont {Demler}, \citenamefont {Jank{\'o}},\ and\ \citenamefont
  {Zar{\'a}nd}}]{moca2008}%
  \BibitemOpen
  \bibfield  {author} {\bibinfo {author} {\bibfnamefont {C.~P.}\ \bibnamefont
  {Moca}}, \bibinfo {author} {\bibfnamefont {E.}~\bibnamefont {Demler}},
  \bibinfo {author} {\bibfnamefont {B.}~\bibnamefont {Jank{\'o}}}, \ and\
  \bibinfo {author} {\bibfnamefont {G.}~\bibnamefont {Zar{\'a}nd}},\ }\bibfield
   {title} {\enquote {\bibinfo {title} {Spin-resolved spectra of {Shiba}
  multiplets from {Mn} impurities in {MgB$_2$}},}\ }\href@noop {} {\bibfield
  {journal} {\bibinfo  {journal} {Phys. Rev. B}\ }\textbf {\bibinfo {volume}
  {77}},\ \bibinfo {pages} {174516} (\bibinfo {year} {2008})}\BibitemShut
  {NoStop}%
\bibitem [{\citenamefont {Balatsky}\ \emph {et~al.}(2006)\citenamefont
  {Balatsky}, \citenamefont {Vekhter},\ and\ \citenamefont
  {Zhu}}]{balatsky2006}%
  \BibitemOpen
  \bibfield  {author} {\bibinfo {author} {\bibfnamefont {A.~V.}\ \bibnamefont
  {Balatsky}}, \bibinfo {author} {\bibfnamefont {I.}~\bibnamefont {Vekhter}}, \
  and\ \bibinfo {author} {\bibfnamefont {Jian-Xin}\ \bibnamefont {Zhu}},\
  }\bibfield  {title} {\enquote {\bibinfo {title} {Impurity-induced states in
  conventional and unconventional superconductors},}\ }\href@noop {} {\bibfield
   {journal} {\bibinfo  {journal} {Rev. Mod. Phys.}\ }\textbf {\bibinfo
  {volume} {78}},\ \bibinfo {pages} {373} (\bibinfo {year} {2006})}\BibitemShut
  {NoStop}%
\bibitem [{\citenamefont {Pillet}\ \emph {et~al.}(2010)\citenamefont {Pillet},
  \citenamefont {Quay}, \citenamefont {Morin}, \citenamefont {Bena},
  \citenamefont {Yeyati},\ and\ \citenamefont {Joyez}}]{pillet2010}%
  \BibitemOpen
  \bibfield  {author} {\bibinfo {author} {\bibfnamefont {J.-D.}\ \bibnamefont
  {Pillet}}, \bibinfo {author} {\bibfnamefont {C.~H.~L.}\ \bibnamefont {Quay}},
  \bibinfo {author} {\bibfnamefont {P.}~\bibnamefont {Morin}}, \bibinfo
  {author} {\bibfnamefont {C.}~\bibnamefont {Bena}}, \bibinfo {author}
  {\bibfnamefont {A.~Levy}\ \bibnamefont {Yeyati}}, \ and\ \bibinfo {author}
  {\bibfnamefont {P.}~\bibnamefont {Joyez}},\ }\bibfield  {title} {\enquote
  {\bibinfo {title} {Andreev bound states in supercurrent-carrying carbon
  nanotubes revealed},}\ }\href@noop {} {\bibfield  {journal} {\bibinfo
  {journal} {Nat. Physics}\ }\textbf {\bibinfo {volume} {6}},\ \bibinfo {pages}
  {965} (\bibinfo {year} {2010})}\BibitemShut {NoStop}%
\bibitem [{\citenamefont {Deacon}\ \emph {et~al.}(2010)\citenamefont {Deacon},
  \citenamefont {Tanaka}, \citenamefont {Oiwa}, \citenamefont {Sakano},
  \citenamefont {Yoshida}, \citenamefont {Shibata}, \citenamefont {Hirakawa},\
  and\ \citenamefont {Tarucha}}]{Deacon:2010jn}%
  \BibitemOpen
  \bibfield  {author} {\bibinfo {author} {\bibfnamefont {R~S}\ \bibnamefont
  {Deacon}}, \bibinfo {author} {\bibfnamefont {Y}~\bibnamefont {Tanaka}},
  \bibinfo {author} {\bibfnamefont {A}~\bibnamefont {Oiwa}}, \bibinfo {author}
  {\bibfnamefont {R}~\bibnamefont {Sakano}}, \bibinfo {author} {\bibfnamefont
  {K}~\bibnamefont {Yoshida}}, \bibinfo {author} {\bibfnamefont
  {K}~\bibnamefont {Shibata}}, \bibinfo {author} {\bibfnamefont
  {K}~\bibnamefont {Hirakawa}}, \ and\ \bibinfo {author} {\bibfnamefont
  {S}~\bibnamefont {Tarucha}},\ }\bibfield  {title} {\enquote {\bibinfo {title}
  {{Interplay of Kondo and superconducting correlations in the nonequilibrium
  Andreev transport through a quantum dot}},}\ }\href@noop {} {\bibfield
  {journal} {\bibinfo  {journal} {Physical Review Letters}\ }\textbf {\bibinfo
  {volume} {104}},\ \bibinfo {pages} {076805} (\bibinfo {year}
  {2010})}\BibitemShut {NoStop}%
\bibitem [{\citenamefont {Maurand}\ \emph {et~al.}(2012)\citenamefont
  {Maurand}, \citenamefont {Meng}, \citenamefont {Bonet}, \citenamefont
  {Florens}, \citenamefont {Marty},\ and\ \citenamefont
  {Wernsdorfer}}]{maurand2012}%
  \BibitemOpen
  \bibfield  {author} {\bibinfo {author} {\bibfnamefont {Romain}\ \bibnamefont
  {Maurand}}, \bibinfo {author} {\bibfnamefont {Tobias}\ \bibnamefont {Meng}},
  \bibinfo {author} {\bibfnamefont {Edgar}\ \bibnamefont {Bonet}}, \bibinfo
  {author} {\bibfnamefont {Serge}\ \bibnamefont {Florens}}, \bibinfo {author}
  {\bibfnamefont {La\"etitia}\ \bibnamefont {Marty}}, \ and\ \bibinfo {author}
  {\bibfnamefont {Wolfgang}\ \bibnamefont {Wernsdorfer}},\ }\bibfield  {title}
  {\enquote {\bibinfo {title} {First-order 0-$\pi$ quantum phase transition in
  the {Kondo} regime of a superconducting carbon-nanotube quantum dot},}\
  }\href@noop {} {\bibfield  {journal} {\bibinfo  {journal} {Phys. Rev. X}\
  }\textbf {\bibinfo {volume} {2}},\ \bibinfo {pages} {011009} (\bibinfo {year}
  {2012})}\BibitemShut {NoStop}%
\bibitem [{\citenamefont {Ji}\ \emph {et~al.}(2008)\citenamefont {Ji},
  \citenamefont {Zhang}, \citenamefont {Fu}, \citenamefont {Chen},
  \citenamefont {Ma}, \citenamefont {Li}, \citenamefont {Duan}, \citenamefont
  {Jia},\ and\ \citenamefont {Xue}}]{ji2008}%
  \BibitemOpen
  \bibfield  {author} {\bibinfo {author} {\bibfnamefont {S.~H.}\ \bibnamefont
  {Ji}}, \bibinfo {author} {\bibfnamefont {T.}~\bibnamefont {Zhang}}, \bibinfo
  {author} {\bibfnamefont {Y.~S.}\ \bibnamefont {Fu}}, \bibinfo {author}
  {\bibfnamefont {X.}~\bibnamefont {Chen}}, \bibinfo {author} {\bibfnamefont
  {Xu-Cun}\ \bibnamefont {Ma}}, \bibinfo {author} {\bibfnamefont
  {J.}~\bibnamefont {Li}}, \bibinfo {author} {\bibfnamefont {Wen-Hui}\
  \bibnamefont {Duan}}, \bibinfo {author} {\bibfnamefont {Jin-Feng}\
  \bibnamefont {Jia}}, \ and\ \bibinfo {author} {\bibfnamefont {Qi-Kun}\
  \bibnamefont {Xue}},\ }\bibfield  {title} {\enquote {\bibinfo {title}
  {High-resolution scanning tunneling spectroscopy of magnetic impurity induced
  bound states in the superconducting gap of {Pb} thin films},}\ }\href@noop {}
  {\bibfield  {journal} {\bibinfo  {journal} {Phys. Rev. Lett.}\ }\textbf
  {\bibinfo {volume} {100}},\ \bibinfo {pages} {226801} (\bibinfo {year}
  {2008})}\BibitemShut {NoStop}%
\bibitem [{\citenamefont {Franke}\ \emph {et~al.}(2011)\citenamefont {Franke},
  \citenamefont {Schulze},\ and\ \citenamefont {Pascual}}]{franke2011}%
  \BibitemOpen
  \bibfield  {author} {\bibinfo {author} {\bibfnamefont {K.~J.}\ \bibnamefont
  {Franke}}, \bibinfo {author} {\bibfnamefont {G.}~\bibnamefont {Schulze}}, \
  and\ \bibinfo {author} {\bibfnamefont {J.~I.}\ \bibnamefont {Pascual}},\
  }\bibfield  {title} {\enquote {\bibinfo {title} {Competition of
  superconductivity phenomena and {Kondo} screening at the nanoscale},}\
  }\href@noop {} {\bibfield  {journal} {\bibinfo  {journal} {Science}\ }\textbf
  {\bibinfo {volume} {332}},\ \bibinfo {pages} {940} (\bibinfo {year}
  {2011})}\BibitemShut {NoStop}%
\bibitem [{\citenamefont {Hatter}\ \emph {et~al.}(2016)\citenamefont {Hatter},
  \citenamefont {Ruby}, \citenamefont {Pascual}, \citenamefont {Heinrich},\
  and\ \citenamefont {Franke}}]{Hatter:2016kg}%
  \BibitemOpen
  \bibfield  {author} {\bibinfo {author} {\bibfnamefont {Nino}\ \bibnamefont
  {Hatter}}, \bibinfo {author} {\bibfnamefont {Michael}\ \bibnamefont {Ruby}},
  \bibinfo {author} {\bibfnamefont {Jos{\'e}~I}\ \bibnamefont {Pascual}},
  \bibinfo {author} {\bibfnamefont {Benjamin~W}\ \bibnamefont {Heinrich}}, \
  and\ \bibinfo {author} {\bibfnamefont {Katharina~J}\ \bibnamefont {Franke}},\
  }\bibfield  {title} {\enquote {\bibinfo {title} {{Magnetic anisotropy in
  Shiba bound states across a quantum phase transition}},}\ }\href@noop {}
  {\bibfield  {journal} {\bibinfo  {journal} {Nature Communications}\ ,\
  \bibinfo {pages} {1--6}} (\bibinfo {year} {2016})}\BibitemShut {NoStop}%
\bibitem [{\citenamefont {Nozi{\`e}res}(1974)}]{nozieres1974}%
  \BibitemOpen
  \bibfield  {author} {\bibinfo {author} {\bibfnamefont {P.}~\bibnamefont
  {Nozi{\`e}res}},\ }\bibfield  {title} {\enquote {\bibinfo {title}
  {Fermi-liquid description of {Kondo} problem at low temperatures},}\
  }\href@noop {} {\bibfield  {journal} {\bibinfo  {journal} {J. Low. Temp.
  Phys.}\ }\textbf {\bibinfo {volume} {17}},\ \bibinfo {pages} {31} (\bibinfo
  {year} {1974})}\BibitemShut {NoStop}%
\bibitem [{\citenamefont {Koller}\ \emph {et~al.}(2005)\citenamefont {Koller},
  \citenamefont {Hewson},\ and\ \citenamefont {Meyer}}]{koller2005}%
  \BibitemOpen
  \bibfield  {author} {\bibinfo {author} {\bibfnamefont {W.}~\bibnamefont
  {Koller}}, \bibinfo {author} {\bibfnamefont {A.~C.}\ \bibnamefont {Hewson}},
  \ and\ \bibinfo {author} {\bibfnamefont {D.}~\bibnamefont {Meyer}},\
  }\bibfield  {title} {\enquote {\bibinfo {title} {Singular dynamics of
  underscreened magnetic impurity models},}\ }\href@noop {} {\bibfield
  {journal} {\bibinfo  {journal} {Phys. Rev. B}\ }\textbf {\bibinfo {volume}
  {72}},\ \bibinfo {pages} {045117} (\bibinfo {year} {2005})}\BibitemShut
  {NoStop}%
\bibitem [{\citenamefont {Mehta}\ \emph {et~al.}(2005)\citenamefont {Mehta},
  \citenamefont {Andrei}, \citenamefont {Coleman}, \citenamefont {Borda},\ and\
  \citenamefont {Zar\'and}}]{mehta2005}%
  \BibitemOpen
  \bibfield  {author} {\bibinfo {author} {\bibfnamefont {P.}~\bibnamefont
  {Mehta}}, \bibinfo {author} {\bibfnamefont {N.}~\bibnamefont {Andrei}},
  \bibinfo {author} {\bibfnamefont {P.}~\bibnamefont {Coleman}}, \bibinfo
  {author} {\bibfnamefont {L.}~\bibnamefont {Borda}}, \ and\ \bibinfo {author}
  {\bibfnamefont {G.}~\bibnamefont {Zar\'and}},\ }\bibfield  {title} {\enquote
  {\bibinfo {title} {Regular and singular {Fermi}-liquid fixed points in
  quantum impurity models},}\ }\href@noop {} {\bibfield  {journal} {\bibinfo
  {journal} {Phys. Rev. B}\ }\textbf {\bibinfo {volume} {72}},\ \bibinfo
  {pages} {014430} (\bibinfo {year} {2005})}\BibitemShut {NoStop}%
\bibitem [{\citenamefont {Hewson}(1993)}]{hewson}%
  \BibitemOpen
  \bibfield  {author} {\bibinfo {author} {\bibfnamefont {A.~C.}\ \bibnamefont
  {Hewson}},\ }\href@noop {} {\emph {\bibinfo {title} {The Kondo Problem to
  Heavy-Fermions}}}\ (\bibinfo  {publisher} {Cambridge University Press,
  Cambridge},\ \bibinfo {year} {1993})\BibitemShut {NoStop}%
\bibitem [{\citenamefont {Mattis}(1968)}]{mattis1968}%
  \BibitemOpen
  \bibfield  {author} {\bibinfo {author} {\bibfnamefont {D.~C.}\ \bibnamefont
  {Mattis}},\ }\bibfield  {title} {\enquote {\bibinfo {title} {Symmetry of
  ground state in a dilute magnetic metal alloy},}\ }\href@noop {} {\bibfield
  {journal} {\bibinfo  {journal} {Phys. Rev. Lett.}\ }\textbf {\bibinfo
  {volume} {19}},\ \bibinfo {pages} {1478} (\bibinfo {year}
  {1968})}\BibitemShut {NoStop}%
\bibitem [{\citenamefont {Nozi{\`e}res}\ and\ \citenamefont
  {Blandin}(1980)}]{nozieres1980}%
  \BibitemOpen
  \bibfield  {author} {\bibinfo {author} {\bibfnamefont {P.}~\bibnamefont
  {Nozi{\`e}res}}\ and\ \bibinfo {author} {\bibfnamefont {A.}~\bibnamefont
  {Blandin}},\ }\bibfield  {title} {\enquote {\bibinfo {title} {Kondo effect in
  real metals},}\ }\href@noop {} {\bibfield  {journal} {\bibinfo  {journal} {J.
  Phys. (Paris)}\ }\textbf {\bibinfo {volume} {41}},\ \bibinfo {pages} {193}
  (\bibinfo {year} {1980})}\BibitemShut {NoStop}%
\bibitem [{\citenamefont {Cragg}\ \emph {et~al.}(1980)\citenamefont {Cragg},
  \citenamefont {Lloyd},\ and\ \citenamefont {Nozi{\`e}res}}]{cragg1980}%
  \BibitemOpen
  \bibfield  {author} {\bibinfo {author} {\bibfnamefont {D.~M.}\ \bibnamefont
  {Cragg}}, \bibinfo {author} {\bibfnamefont {P.}~\bibnamefont {Lloyd}}, \ and\
  \bibinfo {author} {\bibfnamefont {P.}~\bibnamefont {Nozi{\`e}res}},\
  }\bibfield  {title} {\enquote {\bibinfo {title} {On the ground states of some
  s-d exchange {Kondo} hamiltonians},}\ }\href@noop {} {\bibfield  {journal}
  {\bibinfo  {journal} {J. Phys. C: Solid St. Phys.}\ }\textbf {\bibinfo
  {volume} {13}},\ \bibinfo {pages} {803} (\bibinfo {year} {1980})}\BibitemShut
  {NoStop}%
\bibitem [{\citenamefont {Oreg}\ and\ \citenamefont
  {Goldhaber-Gordon}(2003)}]{oreg2003}%
  \BibitemOpen
  \bibfield  {author} {\bibinfo {author} {\bibfnamefont {Y.}~\bibnamefont
  {Oreg}}\ and\ \bibinfo {author} {\bibfnamefont {D.}~\bibnamefont
  {Goldhaber-Gordon}},\ }\bibfield  {title} {\enquote {\bibinfo {title}
  {{Two-channel Kondo effect in a modified single electron transistor}},}\
  }\href@noop {} {\bibfield  {journal} {\bibinfo  {journal} {Phys. Rev. Lett.}\
  }\textbf {\bibinfo {volume} {90}},\ \bibinfo {pages} {136602} (\bibinfo
  {year} {2003})}\BibitemShut {NoStop}%
\bibitem [{\citenamefont {Potok}\ \emph {et~al.}(2007)\citenamefont {Potok},
  \citenamefont {Rau}, \citenamefont {Shtrikman}, \citenamefont {Oreg},\ and\
  \citenamefont {Goldhaber-Gordon}}]{potok2007}%
  \BibitemOpen
  \bibfield  {author} {\bibinfo {author} {\bibfnamefont {R.~M.}\ \bibnamefont
  {Potok}}, \bibinfo {author} {\bibfnamefont {I.~G.}\ \bibnamefont {Rau}},
  \bibinfo {author} {\bibfnamefont {Hadas}\ \bibnamefont {Shtrikman}}, \bibinfo
  {author} {\bibfnamefont {Yuval}\ \bibnamefont {Oreg}}, \ and\ \bibinfo
  {author} {\bibfnamefont {D.}~\bibnamefont {Goldhaber-Gordon}},\ }\bibfield
  {title} {\enquote {\bibinfo {title} {Observation of the two-channel {Kondo}
  effect},}\ }\href@noop {} {\bibfield  {journal} {\bibinfo  {journal}
  {Nature}\ }\textbf {\bibinfo {volume} {446}},\ \bibinfo {pages} {167}
  (\bibinfo {year} {2007})}\BibitemShut {NoStop}%
\bibitem [{\citenamefont {Keller}\ \emph {et~al.}(2015)\citenamefont {Keller},
  \citenamefont {Peeters}, \citenamefont {Moca}, \citenamefont {Weymann},
  \citenamefont {Mahalu}, \citenamefont {Umansky}, \citenamefont {Zar\'and},\
  and\ \citenamefont {Goldhaber-Gordon}}]{keller2015}%
  \BibitemOpen
  \bibfield  {author} {\bibinfo {author} {\bibfnamefont {A.~J.}\ \bibnamefont
  {Keller}}, \bibinfo {author} {\bibfnamefont {L.}~\bibnamefont {Peeters}},
  \bibinfo {author} {\bibfnamefont {C.~P.}\ \bibnamefont {Moca}}, \bibinfo
  {author} {\bibfnamefont {I.}~\bibnamefont {Weymann}}, \bibinfo {author}
  {\bibfnamefont {D.}~\bibnamefont {Mahalu}}, \bibinfo {author} {\bibfnamefont
  {V.}~\bibnamefont {Umansky}}, \bibinfo {author} {\bibfnamefont
  {G.}~\bibnamefont {Zar\'and}}, \ and\ \bibinfo {author} {\bibfnamefont
  {D.}~\bibnamefont {Goldhaber-Gordon}},\ }\bibfield  {title} {\enquote
  {\bibinfo {title} {{Universal Fermi liquid crossover and quantum criticality
  in a mesoscopic system}},}\ }\href@noop {} {\bibfield  {journal} {\bibinfo
  {journal} {Nature}\ }\textbf {\bibinfo {volume} {526}},\ \bibinfo {pages}
  {237} (\bibinfo {year} {2015})}\BibitemShut {NoStop}%
\bibitem [{\citenamefont {Iftikhar}\ \emph {et~al.}(2015)\citenamefont
  {Iftikhar}, \citenamefont {Jezouin}, \citenamefont {Anthore}, \citenamefont
  {Gennser}, \citenamefont {Parmentier}, \citenamefont {Cavanna},\ and\
  \citenamefont {Pierre}}]{iftikhar2015}%
  \BibitemOpen
  \bibfield  {author} {\bibinfo {author} {\bibfnamefont {Z.}~\bibnamefont
  {Iftikhar}}, \bibinfo {author} {\bibfnamefont {S.}~\bibnamefont {Jezouin}},
  \bibinfo {author} {\bibfnamefont {A.}~\bibnamefont {Anthore}}, \bibinfo
  {author} {\bibfnamefont {U.}~\bibnamefont {Gennser}}, \bibinfo {author}
  {\bibfnamefont {F.~D.}\ \bibnamefont {Parmentier}}, \bibinfo {author}
  {\bibfnamefont {A.}~\bibnamefont {Cavanna}}, \ and\ \bibinfo {author}
  {\bibfnamefont {F.}~\bibnamefont {Pierre}},\ }\bibfield  {title} {\enquote
  {\bibinfo {title} {{Two-channel Kondo effect and renormalization flow with
  macroscopic quantum charge states}},}\ }\href@noop {} {\bibfield  {journal}
  {\bibinfo  {journal} {Nature}\ }\textbf {\bibinfo {volume} {526}},\ \bibinfo
  {pages} {233} (\bibinfo {year} {2015})}\BibitemShut {NoStop}%
\bibitem [{\citenamefont {Affleck}(2005)}]{affleck2005}%
  \BibitemOpen
  \bibfield  {author} {\bibinfo {author} {\bibfnamefont {I.}~\bibnamefont
  {Affleck}},\ }\bibfield  {title} {\enquote {\bibinfo {title} {Non-{Fermi}
  liquid behavior in {Kondo} models},}\ }\href@noop {} {\bibfield  {journal}
  {\bibinfo  {journal} {J. Phys. Soc. Japan}\ }\textbf {\bibinfo {volume}
  {74}},\ \bibinfo {pages} {59} (\bibinfo {year} {2005})}\BibitemShut {NoStop}%
\bibitem [{\citenamefont {Affleck}\ and\ \citenamefont
  {Ludwig}(1991{\natexlab{a}})}]{affleck1991over}%
  \BibitemOpen
  \bibfield  {author} {\bibinfo {author} {\bibfnamefont {I.}~\bibnamefont
  {Affleck}}\ and\ \bibinfo {author} {\bibfnamefont {Andreas W.~W.}\
  \bibnamefont {Ludwig}},\ }\bibfield  {title} {\enquote {\bibinfo {title}
  {Critical theory of overscreened {Kondo} fixed points},}\ }\href@noop {}
  {\bibfield  {journal} {\bibinfo  {journal} {Nucl. Phys. B}\ }\textbf
  {\bibinfo {volume} {360}},\ \bibinfo {pages} {641} (\bibinfo {year}
  {1991}{\natexlab{a}})}\BibitemShut {NoStop}%
\bibitem [{\citenamefont {Emery}\ and\ \citenamefont
  {Kivelson}(1992)}]{emery1992}%
  \BibitemOpen
  \bibfield  {author} {\bibinfo {author} {\bibfnamefont {V.~J.}\ \bibnamefont
  {Emery}}\ and\ \bibinfo {author} {\bibfnamefont {S.}~\bibnamefont
  {Kivelson}},\ }\bibfield  {title} {\enquote {\bibinfo {title} {{Mapping of
  the two-channel Kondo problem to a resonant-level model}},}\ }\href@noop {}
  {\bibfield  {journal} {\bibinfo  {journal} {Phys. Rev. B}\ }\textbf {\bibinfo
  {volume} {46}},\ \bibinfo {pages} {10812} (\bibinfo {year}
  {1992})}\BibitemShut {NoStop}%
\bibitem [{\citenamefont {Coleman}\ \emph {et~al.}(1995)\citenamefont
  {Coleman}, \citenamefont {Ioffe},\ and\ \citenamefont
  {Tsvelik}}]{coleman1995}%
  \BibitemOpen
  \bibfield  {author} {\bibinfo {author} {\bibfnamefont {P.}~\bibnamefont
  {Coleman}}, \bibinfo {author} {\bibfnamefont {L.~B.}\ \bibnamefont {Ioffe}},
  \ and\ \bibinfo {author} {\bibfnamefont {A.~M.}\ \bibnamefont {Tsvelik}},\
  }\bibfield  {title} {\enquote {\bibinfo {title} {{Simple formulation of the
  two-channel Kondo model}},}\ }\href@noop {} {\bibfield  {journal} {\bibinfo
  {journal} {Phys. Rev. B}\ }\textbf {\bibinfo {volume} {52}},\ \bibinfo
  {pages} {6611} (\bibinfo {year} {1995})}\BibitemShut {NoStop}%
\bibitem [{\citenamefont {Maldacena}\ and\ \citenamefont
  {Ludwig}(1997)}]{maldacena1997}%
  \BibitemOpen
  \bibfield  {author} {\bibinfo {author} {\bibfnamefont {Juan~M.}\ \bibnamefont
  {Maldacena}}\ and\ \bibinfo {author} {\bibfnamefont {Andreas W.~W.}\
  \bibnamefont {Ludwig}},\ }\bibfield  {title} {\enquote {\bibinfo {title}
  {Majorana fermions, exact mapping between quantum impurity fixed points with
  four bulk fermion species, and solution of the unitarity puzzle},}\
  }\href@noop {} {\bibfield  {journal} {\bibinfo  {journal} {Nucl. Phys. B}\
  }\textbf {\bibinfo {volume} {506}},\ \bibinfo {pages} {565} (\bibinfo {year}
  {1997})}\BibitemShut {NoStop}%
\bibitem [{\citenamefont {Wilson}(1975)}]{wilson1975}%
  \BibitemOpen
  \bibfield  {author} {\bibinfo {author} {\bibfnamefont {K.~G.}\ \bibnamefont
  {Wilson}},\ }\bibfield  {title} {\enquote {\bibinfo {title} {The
  renormalization group: {Critical} phenomena and the {Kondo} problem},}\
  }\href@noop {} {\bibfield  {journal} {\bibinfo  {journal} {Rev. Mod. Phys.}\
  }\textbf {\bibinfo {volume} {47}},\ \bibinfo {pages} {773} (\bibinfo {year}
  {1975})}\BibitemShut {NoStop}%
\bibitem [{\citenamefont {Krishna-murthy}\ \emph {et~al.}(1980)\citenamefont
  {Krishna-murthy}, \citenamefont {Wilkins},\ and\ \citenamefont
  {Wilson}}]{krishna1980a}%
  \BibitemOpen
  \bibfield  {author} {\bibinfo {author} {\bibfnamefont {H.~R.}\ \bibnamefont
  {Krishna-murthy}}, \bibinfo {author} {\bibfnamefont {J.~W.}\ \bibnamefont
  {Wilkins}}, \ and\ \bibinfo {author} {\bibfnamefont {K.~G.}\ \bibnamefont
  {Wilson}},\ }\bibfield  {title} {\enquote {\bibinfo {title}
  {Renormalization-group approach to the {Anderson} model of dilute magnetic
  alloys. {I.} {S}tatic properties for the symmetric case},}\ }\href@noop {}
  {\bibfield  {journal} {\bibinfo  {journal} {Phys. Rev. B}\ }\textbf {\bibinfo
  {volume} {21}},\ \bibinfo {pages} {1003} (\bibinfo {year}
  {1980})}\BibitemShut {NoStop}%
\bibitem [{\citenamefont {Bulla}\ \emph {et~al.}(2008)\citenamefont {Bulla},
  \citenamefont {Costi},\ and\ \citenamefont {Pruschke}}]{bulla2008}%
  \BibitemOpen
  \bibfield  {author} {\bibinfo {author} {\bibfnamefont {Ralf}\ \bibnamefont
  {Bulla}}, \bibinfo {author} {\bibfnamefont {Theo}\ \bibnamefont {Costi}}, \
  and\ \bibinfo {author} {\bibfnamefont {Thomas}\ \bibnamefont {Pruschke}},\
  }\bibfield  {title} {\enquote {\bibinfo {title} {The numerical
  renormalization group method for quantum impurity systems},}\ }\href@noop {}
  {\bibfield  {journal} {\bibinfo  {journal} {Rev. Mod. Phys.}\ }\textbf
  {\bibinfo {volume} {80}},\ \bibinfo {pages} {395} (\bibinfo {year}
  {2008})}\BibitemShut {NoStop}%
\bibitem [{\citenamefont {Satori}\ \emph {et~al.}(1992)\citenamefont {Satori},
  \citenamefont {Shiba}, \citenamefont {Sakai},\ and\ \citenamefont
  {Shimizu}}]{satori1992}%
  \BibitemOpen
  \bibfield  {author} {\bibinfo {author} {\bibfnamefont {Koji}\ \bibnamefont
  {Satori}}, \bibinfo {author} {\bibfnamefont {Hiroyuki}\ \bibnamefont
  {Shiba}}, \bibinfo {author} {\bibfnamefont {Osamu}\ \bibnamefont {Sakai}}, \
  and\ \bibinfo {author} {\bibfnamefont {Yukihiro}\ \bibnamefont {Shimizu}},\
  }\bibfield  {title} {\enquote {\bibinfo {title} {Numerical renormalization
  group study of magnetic impurities in superconductors},}\ }\href@noop {}
  {\bibfield  {journal} {\bibinfo  {journal} {J. Phys. Soc. Japan}\ }\textbf
  {\bibinfo {volume} {61}},\ \bibinfo {pages} {3239} (\bibinfo {year}
  {1992})}\BibitemShut {NoStop}%
\bibitem [{\citenamefont {Cox}\ and\ \citenamefont
  {Zawadowski}(1998)}]{cox1998}%
  \BibitemOpen
  \bibfield  {author} {\bibinfo {author} {\bibfnamefont {D.~L.}\ \bibnamefont
  {Cox}}\ and\ \bibinfo {author} {\bibfnamefont {A.}~\bibnamefont
  {Zawadowski}},\ }\bibfield  {title} {\enquote {\bibinfo {title} {Exotic
  {Kondo} effects in metals: magnetic ions in a crystalline electric field and
  tunneling centres},}\ }\href@noop {} {\bibfield  {journal} {\bibinfo
  {journal} {Adv. Phys.}\ }\textbf {\bibinfo {volume} {47}},\ \bibinfo {pages}
  {599} (\bibinfo {year} {1998})}\BibitemShut {NoStop}%
\bibitem [{\citenamefont {Pang}\ and\ \citenamefont {Cox}(1991)}]{pang1991}%
  \BibitemOpen
  \bibfield  {author} {\bibinfo {author} {\bibfnamefont {H.~B.}\ \bibnamefont
  {Pang}}\ and\ \bibinfo {author} {\bibfnamefont {D.~L.}\ \bibnamefont {Cox}},\
  }\bibfield  {title} {\enquote {\bibinfo {title} {Stability of the fixed point
  of the two-channel {Kondo} hamiltonian},}\ }\href@noop {} {\bibfield
  {journal} {\bibinfo  {journal} {Phys. Rev. B}\ }\textbf {\bibinfo {volume}
  {44}},\ \bibinfo {pages} {9454} (\bibinfo {year} {1991})}\BibitemShut
  {NoStop}%
\bibitem [{\citenamefont {Kolf}\ and\ \citenamefont {Kroha}(2007)}]{kolf2007}%
  \BibitemOpen
  \bibfield  {author} {\bibinfo {author} {\bibfnamefont {Christian}\
  \bibnamefont {Kolf}}\ and\ \bibinfo {author} {\bibfnamefont {Johann}\
  \bibnamefont {Kroha}},\ }\bibfield  {title} {\enquote {\bibinfo {title}
  {{Strong versus weak coupling duality and coupling dependence of the Kondo
  temperature in the two-channel Kondo model}},}\ }\href@noop {} {\bibfield
  {journal} {\bibinfo  {journal} {Phys. Rev. B}\ }\textbf {\bibinfo {volume}
  {75}},\ \bibinfo {pages} {045129} (\bibinfo {year} {2007})}\BibitemShut
  {NoStop}%
\bibitem [{sup()}]{supplscnfl}%
  \BibitemOpen
  \href@noop {} {}\bibinfo {note} {The Supplemental Materials contain
  additional results for the 2CK model, the sub-gap spectra of the
  three-channel Kondo model and their interpretation in terms of the effective
  zero-bandwidth model, the conformal field theory discussion of the effect of
  the pairing operator, and a semi-quantitative calculation of the bound states
  within the Anderson-Yuval picture}\BibitemShut {NoStop}%
\bibitem [{\citenamefont {Hecht}\ \emph {et~al.}(2008)\citenamefont {Hecht},
  \citenamefont {Weichselbaum}, \citenamefont {von Delft},\ and\ \citenamefont
  {Bulla}}]{hecht2008}%
  \BibitemOpen
  \bibfield  {author} {\bibinfo {author} {\bibfnamefont {T.}~\bibnamefont
  {Hecht}}, \bibinfo {author} {\bibfnamefont {A.}~\bibnamefont {Weichselbaum}},
  \bibinfo {author} {\bibfnamefont {J.}~\bibnamefont {von Delft}}, \ and\
  \bibinfo {author} {\bibfnamefont {R.}~\bibnamefont {Bulla}},\ }\bibfield
  {title} {\enquote {\bibinfo {title} {Numerical renormalization group
  calculation of near-gap peaks in spectral functions of the {Anderson} model
  with superconducting leads},}\ }\href@noop {} {\bibfield  {journal} {\bibinfo
   {journal} {J. Phys. Condens. Mat.}\ }\textbf {\bibinfo {volume} {20}},\
  \bibinfo {pages} {275213} (\bibinfo {year} {2008})}\BibitemShut {NoStop}%
\bibitem [{\citenamefont {Affleck}\ and\ \citenamefont
  {Ludwig}(1991{\natexlab{b}})}]{affleck1991prl}%
  \BibitemOpen
  \bibfield  {author} {\bibinfo {author} {\bibfnamefont {Ian}\ \bibnamefont
  {Affleck}}\ and\ \bibinfo {author} {\bibfnamefont {Andreas W.~W.}\
  \bibnamefont {Ludwig}},\ }\bibfield  {title} {\enquote {\bibinfo {title}
  {Universal noninteger ground-state degeneracy in critical quantum systems},}\
  }\href@noop {} {\bibfield  {journal} {\bibinfo  {journal} {Phys. Rev. Lett.}\
  }\textbf {\bibinfo {volume} {67}},\ \bibinfo {pages} {161} (\bibinfo {year}
  {1991}{\natexlab{b}})}\BibitemShut {NoStop}%
\bibitem [{\citenamefont {Yuval}\ and\ \citenamefont
  {Anderson}(1970)}]{Anderson&Yuval}%
  \BibitemOpen
  \bibfield  {author} {\bibinfo {author} {\bibfnamefont {G.}~\bibnamefont
  {Yuval}}\ and\ \bibinfo {author} {\bibfnamefont {P.~W.}\ \bibnamefont
  {Anderson}},\ }\bibfield  {title} {\enquote {\bibinfo {title} {Exact results
  for the kondo problem: One-body theory and extension to finite
  temperature},}\ }\href {\doibase 10.1103/PhysRevB.1.1522} {\bibfield
  {journal} {\bibinfo  {journal} {Phys. Rev. B}\ }\textbf {\bibinfo {volume}
  {1}},\ \bibinfo {pages} {1522--1528} (\bibinfo {year} {1970})}\BibitemShut
  {NoStop}%
\bibitem [{\citenamefont {Anderson}\ \emph {et~al.}(1970)\citenamefont
  {Anderson}, \citenamefont {Yuval},\ and\ \citenamefont
  {Hamann}}]{Anderson&Yuval&Hamann}%
  \BibitemOpen
  \bibfield  {author} {\bibinfo {author} {\bibfnamefont {P.~W.}\ \bibnamefont
  {Anderson}}, \bibinfo {author} {\bibfnamefont {G.}~\bibnamefont {Yuval}}, \
  and\ \bibinfo {author} {\bibfnamefont {D.~R.}\ \bibnamefont {Hamann}},\
  }\bibfield  {title} {\enquote {\bibinfo {title} {Exact results in the kondo
  problem. ii. scaling theory, qualitatively correct solution, and some new
  results on one-dimensional classical statistical models},}\ }\href {\doibase
  10.1103/PhysRevB.1.4464} {\bibfield  {journal} {\bibinfo  {journal} {Phys.
  Rev. B}\ }\textbf {\bibinfo {volume} {1}},\ \bibinfo {pages} {4464--4473}
  (\bibinfo {year} {1970})}\BibitemShut {NoStop}%
\bibitem [{\citenamefont {Fabrizio}\ \emph {et~al.}(1995)\citenamefont
  {Fabrizio}, \citenamefont {Gogolin},\ and\ \citenamefont
  {Nozi{\`e}res}}]{1995PhRvL..74.4503F}%
  \BibitemOpen
  \bibfield  {author} {\bibinfo {author} {\bibfnamefont {M}~\bibnamefont
  {Fabrizio}}, \bibinfo {author} {\bibfnamefont {AO}~\bibnamefont {Gogolin}}, \
  and\ \bibinfo {author} {\bibfnamefont {P}~\bibnamefont {Nozi{\`e}res}},\
  }\bibfield  {title} {\enquote {\bibinfo {title} {{Crossover from
  non-Fermi-liquid to Fermi-liquid behavior in the two channel Kondo model with
  channel anisotropy}},}\ }\href@noop {} {\bibfield  {journal} {\bibinfo
  {journal} {Physical Review Letters}\ }\textbf {\bibinfo {volume} {74}},\
  \bibinfo {pages} {4503--4506} (\bibinfo {year} {1995})}\BibitemShut {NoStop}%
\end{thebibliography}%

\end{document}